%% file: KOI205_v5b.tex
\documentclass[letter, traditabstract]{aa}
\pdfoutput=1
\usepackage{hyperref}
\usepackage{txfonts}
\usepackage{graphicx}
\usepackage{natbib}
\usepackage{epsf}

\def \sophie{SOPHIE}
\def \kepler{\emph{Kepler}}

\def\kepler{\emph{Kepler}}
\def \MJ{M$_{\mathrm{Jup}}$}
\def \RJ{R$_{\mathrm{Jup}}$}
\def \RS{R$_{\odot}$}
\def \msol{M$\mathrm{_\odot}$}
\def \teff{$T_\mathrm{eff}$}
\def \kms{km\,s$^{-1}$}
\def \ms{m\,s$^{-1}$}
\def \1s{$1\,\sigma$}

\def \t0{T$_0$}

\def \205{KOI-205}

\begin{document}

\title{SOPHIE velocimetry of \kepler\ transit candidates \thanks{Based on observations collected with the {\it SOPHIE}  spectrograph on the 1.93-m telescope at Observatoire de Haute-Provence (CNRS), France, and with the ESPaDOnS spectrograph on the CFH telescope.}}
            
\subtitle{VIII. KOI-205 b: a brown-dwarf companion to a K-type dwarf.}

\author{
R.~F.~D\'iaz\inst{1}, C.~Damiani\inst{1}, M.~Deleuil\inst{1}, J.~M.~Almenara\inst{1}, C.~Moutou\inst{1}, S.~C.~C.~Barros\inst{1}, A.~S.~Bonomo\inst{1,4}, F.~Bouchy\inst{2,3}, G.~Bruno\inst{1}, G.~H\'ebrard\inst{2,3}, G.~Montagnier\inst{2,3},  A.~Santerne\inst{1}
}

\institute{
Aix Marseille Universit\'e, CNRS, LAM (Laboratoire d'Astrophysique de Marseille) UMR 7326, 13388, Marseille, France \and 
Institut d'Astrophysique de Paris, UMR7095 CNRS, Universit\'e Pierre \& Marie Curie, 98bis boulevard Arago, 75014 Paris, France \and 
Observatoire de Haute-Provence, CNRS/OAMP, 04870 Saint-Michel-l'Observatoire, France \and
INAF - Osservatorio Astronomico di Torino, via Osservatorio 20, 10025, Pino Torinese, Italy
}

 \date{Received TBC; accepted TBC}
      
\abstract{We report the discovery of a transiting brown dwarf companion to \205, a K0 main-sequence star, in a 11.720125-day period orbit. The transits were detected by the \kepler\ space telescope, and the reflex motion of the star was measured using radial velocity observations obtained with the SOPHIE spectrograph. The atmospheric parameters of the host stars were determined from the analysis of high-resolution, high signal-to-noise ratio ESPaDOns spectra obtained for this purpose. Together with spectrophotometric measurements recovered from the literature, these spectra indicate that the star is a mildly metallic K0 dwarf with \teff\ $5237 \pm 60$ K. The mass of the companion is $39.9 \pm 1.0$ \MJ\ and its radius is $0.81 \pm 0.02$ \RJ, in agreement with current theoretical predictions. This is the first time a \emph{bona fide} brown dwarf companion is detected in orbit around a star of this type. The formation and orbital evolution of brown dwarf companions is briefly discussed in the light of this new discovery.}

\authorrunning{D\'iaz et al.}
\titlerunning{KOI-205 b: a brown dwarf companion to a K-type dwarf.}


\maketitle

\section{Introduction}
The formation and evolution of brown dwarfs (BD) is not well understood. While it seems clear that these objects are too massive to generally form by core-accretion in a circumstellar disk \citep[Fig.~3 of][]{mordasini2009}, a number of plausible formation theories have been advanced \citep[see reviews by][]{whitworth2007, luhman2012}, but the role each of them plays in the actual formation process is not clear. The small number of BD companions to solar-type stars prevents reliable statistical studies from being made, and although radial-velocity surveys have detected a few tens of objects with minimum masses in the BD regime \citep[e.g.][]{sahlmann2011, diaz2012}, the inclination degeneracy does not allow us to be certain that these objects are actually sub-stellar. Since the mass function is expected to rise in the stellar regime, the inclination effect is more severe for BD companions \citep{hoturner2011}. On the other hand, in recent years transiting surveys have provided a handful of \emph{bona fide} BD companions \citep[e.g.][]{bouchy2011b, anderson2011}.

Most of the discovered short-period BD companions are orbiting around F-type stars. This led \citet{bouchy2011} to advance the hypothesis that BD companions could not survive in close orbit around stars with convective envelopes ($T_\mathrm{eff}$ between 5000 K and 6000 K), since the tides excited on them by the massive companion and the progressive spin-down due to magnetic braking would lead to the reduction of the orbit size and to the eventual collision with the host star. Because F-type stars are less affected by braking throughout their lives, \citet{bouchy2011} suggested they might be the most common long-term hosts for massive companions.

In this letter we report the discovery of the first short-period BD orbiting a K-type dwarf, \205, and discuss its implications for BD formation and evolution. This object was part of the SOPHIE follow-up of \kepler\ candidates described by \citet{santerne2012}. Like other objects followed-up with SOPHIE, \205 shows that a significant fraction of the \kepler\ Objects of Interest are not transiting planets.

\section{Description of the data}
\subsection{\kepler\ photometry}
\205 (KIC 7046804) was observed by \kepler\ from Quarters 1 through 13, i.e., between May 2009 and  June 2012, with a sampling of one point every 29.4 minutes (Long Cadence data, LC). Although Short Cadence data are also available for some quarters, only LC data were used to reduce computation time, and because the transit shape is sampled well enough in the phase-folded data (Fig.~\ref{fig.results205}). The light curve issued from the Photometric Analysis module of the \kepler\ pipeline was used. One-percent-deep transits are clearly visible by eye; they occur every 11.7 days. The typical uncertainty in individual LC points is around 230 ppm.

For the transit modeling only fragments of the light curve around each transit were considered. Each fragment was normalized with a parabolic fit to the out-of-transit part, and a sigma-clipping at 3$\sigma$ was performed to reject outliers. The contamination by nearby stars was corrected using the crowding values from the \kepler\ archive\footnote{\url{http://archive.stsci.edu/kepler/kepler_fov/search.php}}. Due to the quarterly rotation of the spacecraft, the crowding value changes every three months, but recurs every four seasons. After contamination correction, it is obvious that transits from Season 2 are systematically deeper. We believe this is caused by an overestimation of the crowding factor, which is 12\% in Season 2 and between 6 and 7 \% in the other seasons. We therefore decided to analyze the light curves from quarters belonging to Season 2 separately, and fit a contamination factor independently.

\subsection{SOPHIE velocimetry}
Seven spectra of \205 were obtained between February 2012 and June 2012 with the SOPHIE spectrograph \citep{perruchot2008, bouchy2013}. Observations were performed in high-efficiency mode, with resolving power $\lambda/\Delta\lambda\sim 40\,000$, and exposure times ranging from 900 to 2700 seconds. The resulting signal-to-noise ratios (S/N) per pixel at 550 nm are between 7 and 16.  

The spectra were reduced and extracted using the \sophie\ pipeline \citep{bouchy2009}, and the radial velocites were obtained from a Gaussian fit to the cross-correlation function with numerical masks corresponding to different spectral types. The charge transfer inefficiency (CTI) effect was corrected using the polynomial in \citet{santerne2012}. For faint targets such as these, the spectral orders at the edge of the wavelength range are usually too noisy, and adding them in the average cross-correlation function degrades the precision of the measurements. For \205, 8 and 5 orders were discarded from the blue and red ends of the spectrum, respectively. The resulting radial velocities are plotted in Fig.~\ref{fig.results205} and listed in Table~\ref{table.rv}.

No bisector effect \citep{queloz2001} or mask effect were detected in the SOPHIE data, supporting the sub-stellar companion hypothesis. Bisector measurements are also given in Table~\ref{table.rv}.

%\onltab{
\begin{table}
\caption{Target coordinates and apparent magnitudes.  \label{table.obslog}}
\setlength{\tabcolsep}{5pt}
\begin{tabular}{llll}
\hline
\hline
\noalign{\smallskip}
\multicolumn{2}{r}{\kepler\ ID} 		& \multicolumn{2}{l}{7046804}\\
\multicolumn{2}{r}{2MASS ID}	& \multicolumn{2}{l}{19415919+4232163}\\
R.A. (J2000) & 19 41 59.20 & Dec. (J2000) & +42\degr 32\arcmin 16\farcs4\\
\noalign{\smallskip}
\noalign{\smallskip}
\kepler\ mag\tablefootmark{a} 		& 14.518   & 2MASS-J \tablefootmark{b} 			& $13.095\pm0.021$\\
SDSS $g^\prime$ \tablefootmark{a}		& $15.19\pm0.04$ & 2MASS-H \tablefootmark{b} 			& $12.709\pm0.024$\\
SDSS $r^\prime$ \tablefootmark{a}		& $14.47\pm0.04$ & 2MASS-Ks \tablefootmark{b} 			& $12.656\pm0.021$\\
SDSS $i^\prime$ \tablefootmark{a}		& $14.23\pm0.04$ & WISE-W1 \tablefootmark{c} 			& $12.564\pm0.023$\\
SDSS $z^\prime$ \tablefootmark{a}		& $14.12\pm0.04$ & WISE-W2 \tablefootmark{c} 			& $12.688\pm0.026$\\
\noalign{\smallskip}
\hline
\hline
\end{tabular}
\tablefoot{
\tablefoottext{a}{From the \kepler\ Input Catalogue.}
\tablefoottext{b}{\citet{2MASS}}
\tablefoottext{c}{see \citet{cutri2012}}
}
\end{table}
%}

\section{Host star \label{sec.stellarparams}}
The stellar atmospheric parameters were obtained using eight ESPaDOns spectra, acquired in service mode using the star + sky configuration, which provides a spectral resolution of 65,000 on the night of June 30 2012 (program 12AF05). The resulting combined spectrum has a  S/N $\sim$ 90 at 550 nm per resolution element. The method is described in \citet{deleuil2012} and yields the values $T_\mathrm{eff} = 5210 \pm 70$ K, [Fe/H] $= 0.27 \pm 0.14$ dex, and $\log g = 4.65 \pm 0.07$ [cgs]. The projected rotational velocity was determined to be $v \sin i_* = 2 \pm 1$ \kms. A comparison with StarEvol evolution tracks \citep{lagarde2012} indicates that the star is a K0 dwarf, with a wide range of allowed ages, because of the slow evolution of low-mass stars. However, its slow rotation suggests that the star is older than the lower end of the range given in Table~\ref{table.Params}.

\section{Modeling of the data and parameter estimation \label{sect.modeling}}
The \kepler\ light curve and SOPHIE radial velocities were fitted together with the photometric measurements from Table~\ref{table.obslog}, which provide a well-sampled spectral energy distribution (SED),  to a model consisting of a star orbited by a dark companion of a given mass and radius. The model and the Markov Chain Monte Carlo algorithm used to take samples from the Bayesian posterior are implemented in the PASTIS package, which is described in detail in D\'iaz et al.\ (2013, in prep.). Basically, we modeled the light curves with the EBOP code \citep{etzel1981, popperetzel1981} using a quadratic limb-darkening law with coefficients interpolated from the tables of \citet{claret2011}. To deal with the distortion of the transit shape arising from the finite integration time \citep{kipping2010} of LC data, we oversampled the model to five times the original sampling rate and re-binned to the LC sampling rate before comparing them to the data. The radial velocities were fitted to a Keplerian orbit. The model SED was obtained using an interpolation of the PHOENIX/BT-Settl synthetic spectral library \citep{allard2012} for a given \teff, [Fe/H], and $\log g$ of the host star. The interpolated spectrum is scaled to a given distance and corrected from interstellar extinction. The distance $d$ and the color excess $E(B-V)$ were included as free parameters in our model.

The model therefore has 14 parameters, which are marked in Table~\ref{table.Params}. The mass ratio $q$ has only a slight effect on the model and because of the normalization procedure of the light curve described above, it is not constrained by these data. It was nevertheless included as a free parameter to take into account its effects on the error budget of the remaining parameters. Additionally, we included parameters describing the data: the out-of-transit flux of each of the two \kepler\ light curves, and a contamination factor for Season 2. Finally, the systematic errors in the radial velocities and in the light curves are modeled as an additional source of Gaussian noise with zero mean and variance $\sigma^2_J$. The widths $\sigma_J$ of each dataset were also fitted. 

Samples from the joint posterior distribution of the parameters were obtained using the Metropolis-Hastings algorithm \citep[e.g.][]{tegmark2004} with an adaptive step size \citep{ford2006}, coupled with an adaptive principal component analysis to correctly sample the posterior {\bf even in the presence of  non-linear correlations}. Non-informative priors --i.e., uniform or Jeffreys distributions-- were used for all parameters except for \teff, $z$, $P$, and $T_c$, for which a normal distribution was deemed more appropriate. For $P$, and $T_c$ the determination by \citet{batalha2012} was used, but the width of the distribution was increased by an order of magnitude to ensure that it would not bias the results.

\begin{figure}[t]
\begin{center}
\includegraphics[width=0.7\columnwidth]{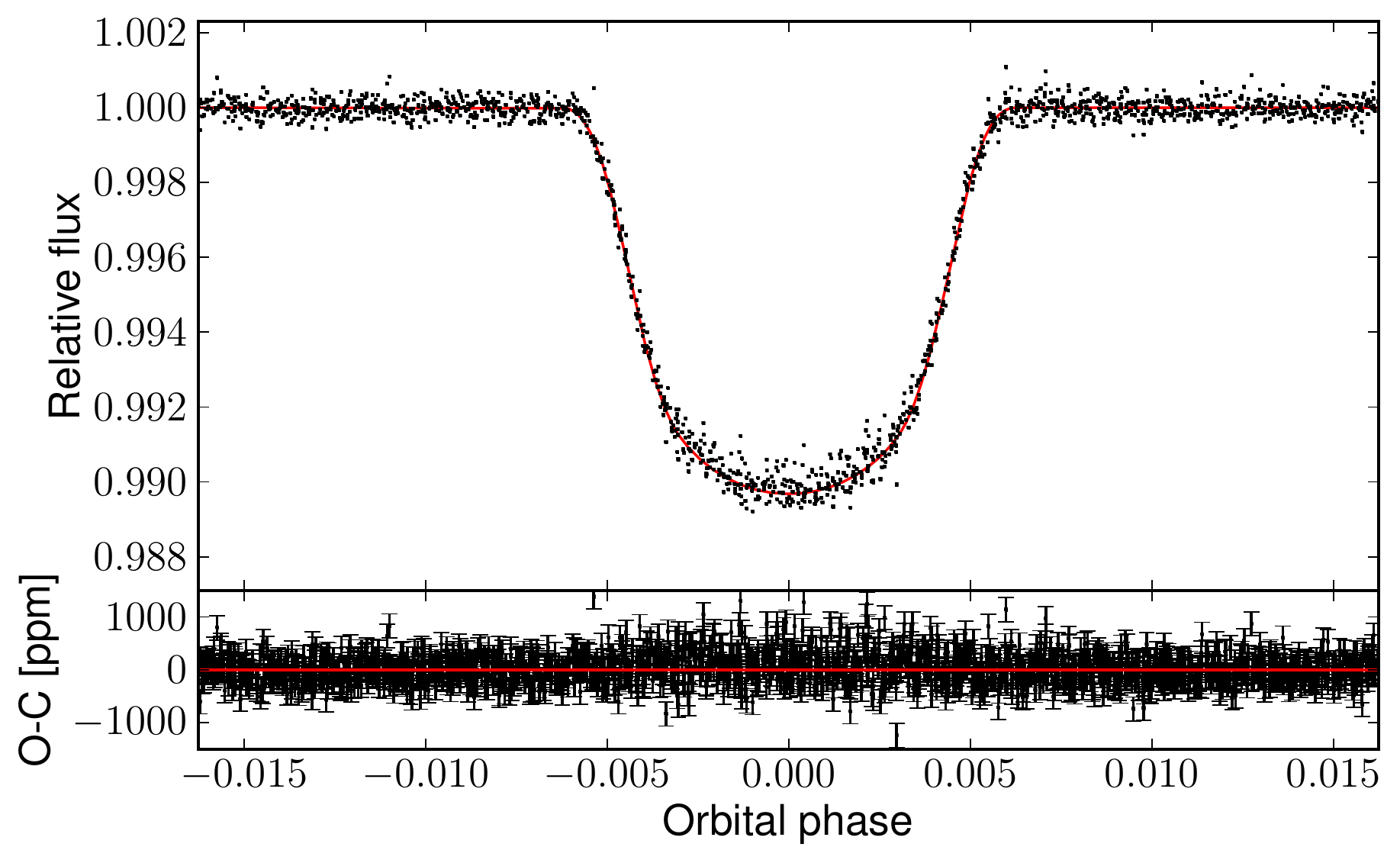}\hfill
\includegraphics[width=0.7\columnwidth]{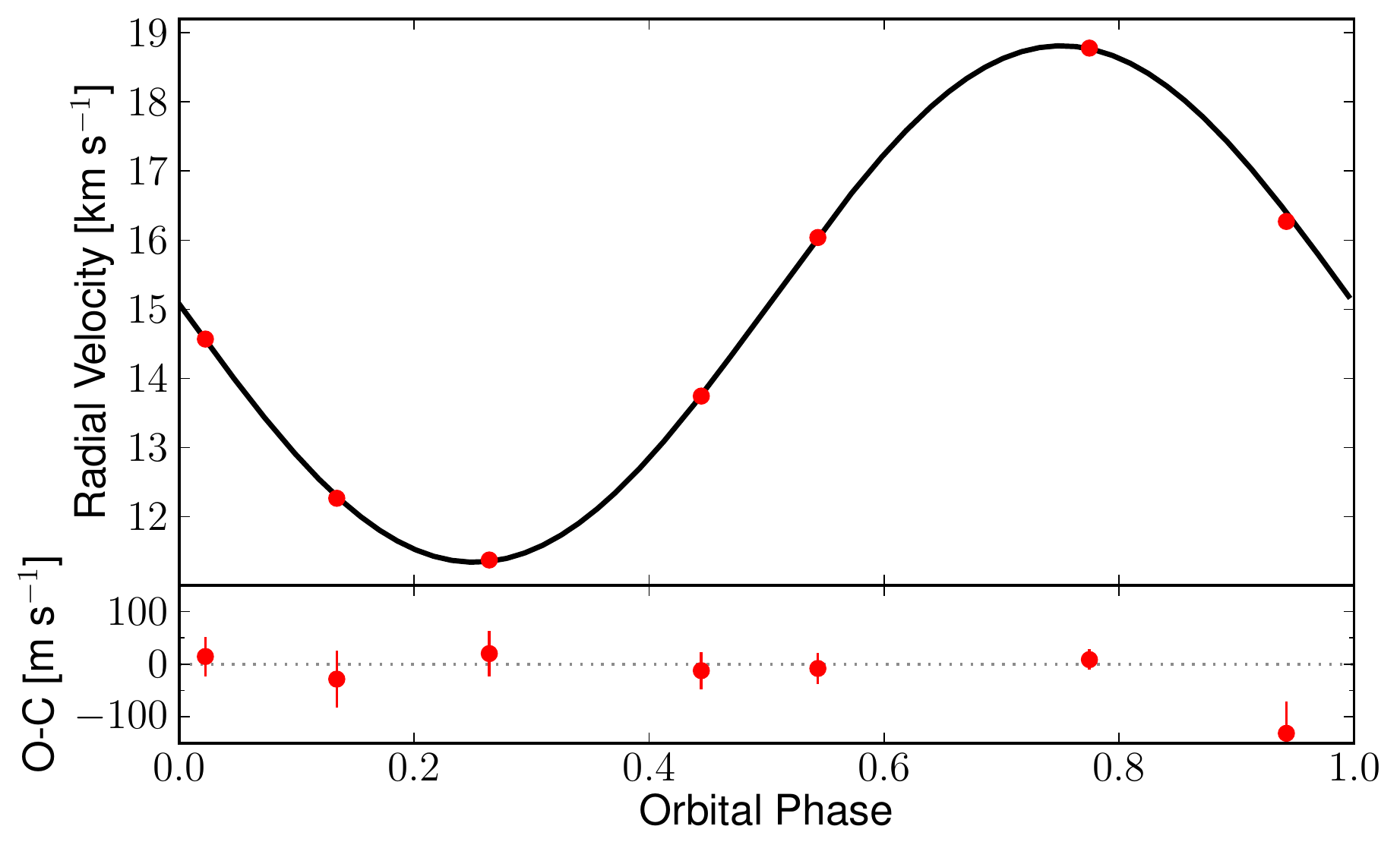}\hfill
\includegraphics[width=0.7\columnwidth]{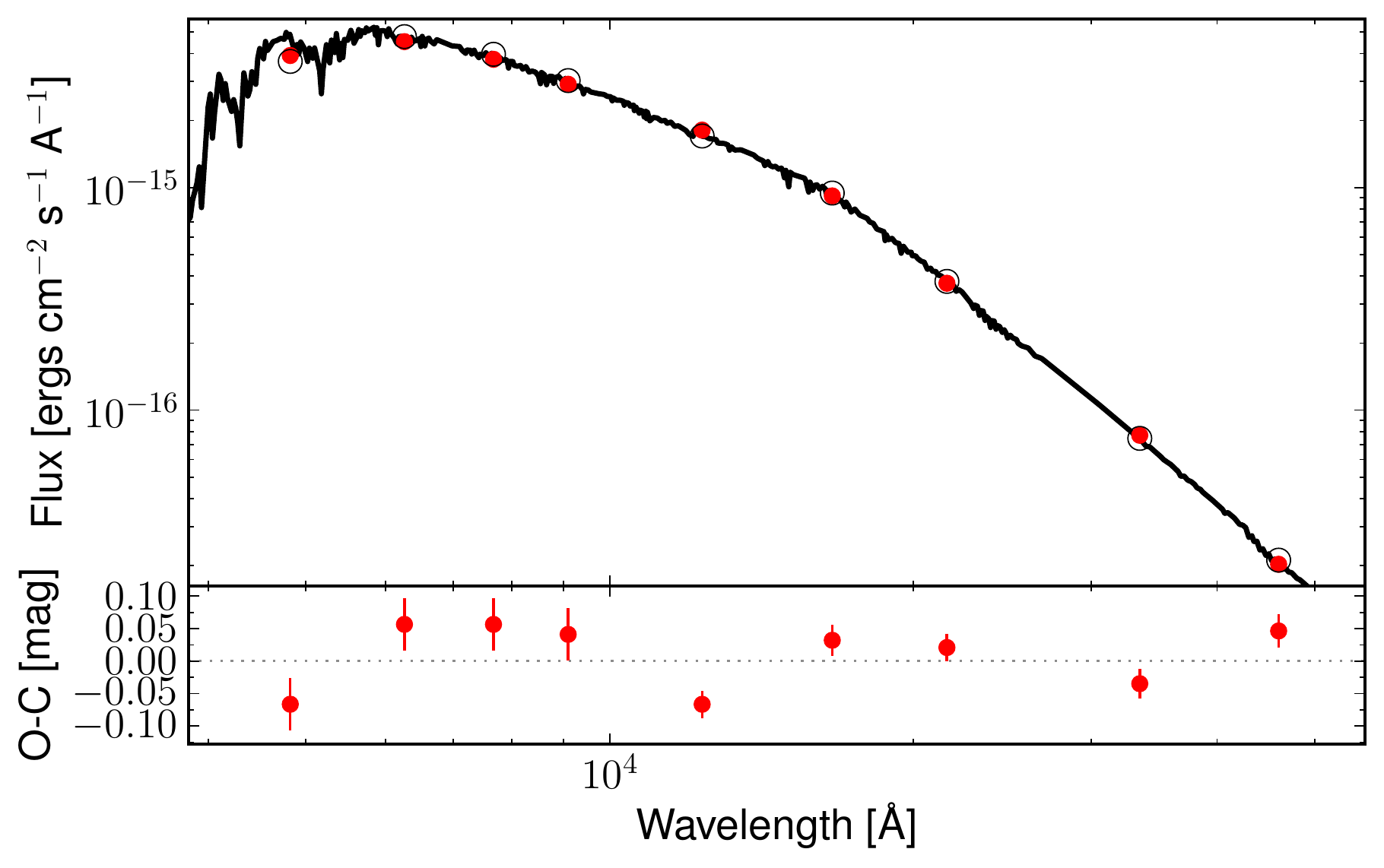}
\caption{Data and best-fit model for KOI-205. For the SED, the best-fit spectrum is plotted as a solid black curve, and the integrated flux in each of the photometric bands is plotted as open circles. \label{fig.results205}}
\end{center}
\end{figure}

Forty chains of 700,000 steps each were started at random points drawn from the joint prior, and 25 of them appeared to converge to the same solution. The remaining ones were stuck in regions of lower posterior probability, and were not considered further. A modified version \citet{geweke1992} diagnostic was used to determine when each chain had reached a stationary state. To ensure that the  samples used to estimate the parameters and their uncertainties are independent, the chains were thinned using their correlation length \citep[e.g.][]{tegmark2004} before merging. In total, over 20,000 independent samples of the posterior distribution were obtained in this way. The formal 68.3\% confidence intervals are listed in Table~\ref{table.Params} and Fig.~\ref{fig.results205} presents the maximum posterior model. The median value for $\sigma_J$ was found to be roughly 100 ppm for \kepler\ light curves\footnote{This value seems much higher than the typical value reported by \citet{gilliland2011}, but the noise metric they used greatly differs from ours. A rough transformation of $\sigma_J$ to their metric yields a value that agrees with the stellar noise distribution presented in that paper.}, and around 40 \ms\ for SOPHIE velocities, possibly due to the low S/N of the spectra and to incomplete CTI correction. The light curve contamination of Season 2 is around 6\%.

%\onltab{
\begin{table*}[t]
\caption{Radial velocity measurements for \205 \label{table.rv}}
\begin{tabular}{l l l c c c}
\hline
\hline
\noalign{\smallskip}
BJD     & RV          & $\sigma_{RV}$ & BVS\tablefootmark{a}  & Exp.\ time & S/N/pix    \\
-2 450 000  &(\kms) & (\kms) & (\kms) & (s)       & (550 nm)\\
\hline
\noalign{\smallskip}
5984.6759 &12.144 & 0.054 &  -0.386 & 1800 &  10\\
6015.6232 &18.724 & 0.019 &  -0.077 & 2700  & 16\\
6064.4682 &16.055 & 0.061 &   0.193  & 1200 &  7\\
6071.5153 &15.932 & 0.030 &   0.061  &1371 & 11\\
6100.5669 &14.414 & 0.038 & -0.035  &1079 &  9\\
6103.3985 &11.213 & 0.043 & -0.052 & 900   &   9\\
6105.5132 &13.640 & 0.036 &  0.017  &1200 & 11\\
\noalign{\smallskip}
\hline 
\end{tabular}
\tablefoot{
\tablefoottext{a}{Bisector velocity span.}
}
\end{table*}
%}

\section{Results and discussion}
\begin{figure}[t!]
\begin{center}
\includegraphics[width=\columnwidth]{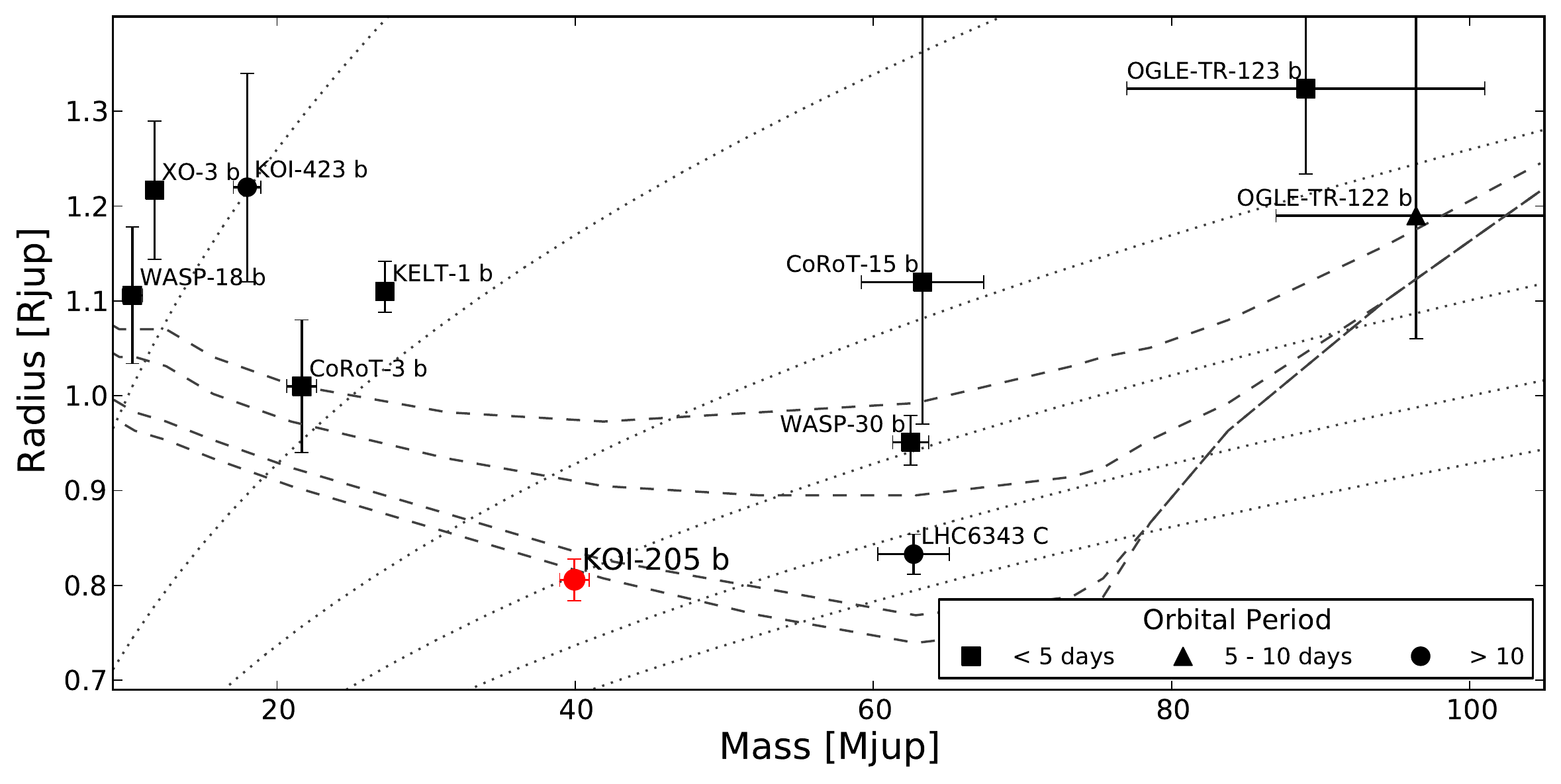}
\caption{Mass radius diagram including transiting companions more massive than 10 \MJ. The dashed curves are the \citet{baraffe2003} isochrones for (from top to bottom) 0.5, 1, 5, and 10 gigayears.  The dotted lines are the isodensity curves for 10, 25, 50, 75, 100, and 125 times the mean density of Jupiter. \citep[Data from:][]{siverd2012, triaud2013, bouchy2011, bouchy2011b, johnson2011, hellier2009, deleuil2008, winn2008, pont2005a, pont2006c} \label{fig.mass_radius}}
\end{center}
\end{figure}

Our analysis shows that \205 has a BD companion with a mass of 40 \MJ\ and a radius of 0.8 \RJ, in a 11.7-day circular orbit. The position of \205 b in the mass-radius diagram (Fig.~\ref{fig.mass_radius}) agrees well with the theoretical isochrones by \citet{baraffe2003} for a system with an age between 5 and 10 Gyr. If the age of the system were closer to 1 Gyr, which is permitted by the data, then the models would not reproduce the observed radius. However, given the slow rotation rate of \205, it is likely that the age of the system is closer to 5 Gyr than to the lower end of the range reported in table~\ref{table.Params}. Indeed, an analysis of the pipeline-corrected PDC \kepler\ light curve yields a rotational period of around 41 days, in agreement with the spectroscopic $v \sin i_*$ value. Although \205 b is the smallest BD detected as yet, its bulk density is similar to that of \object{WASP-30} b \citep{triaud2013, anderson2011} and lower than that of \object{LHC6343C} \citep{johnson2011}.

On the other hand, \205 b is remarkable in two senses. First, it is the only object in the mass regime between massive BDs, such as \object{WASP-30} b or \object{CoRoT-15} b, and light BDs (or massive planets) such as \object{KOI-423} b and \object{CoRoT-3} b. Secondly, it is the only short-period ($P \lesssim 10-15$ days) transiting BD known to date in orbit around a K-type star. Indeed, the second-most massive object orbiting around a similar star is \object{HAT-P-20} b  \citep{bakos2011}, with a mass of 7 \MJ,  i.e., almost six times less massive. Moreover, radial velocity surveys have discovered only four non-transiting objects with minimum masses above 10 \MJ\ orbiting stars less massive than 1 \msol\ in short-period orbits. Given that observational biases tend to favor the detection of this population, its apparent paucity raises questions about its origin.

Formation of BDs can proceed by gravitational collapse of a molecular cloud in a similar fashion as stellar objects form. Recent simulations \citep{bate2012} have shown that BDs can be formed in binaries with stellar primaries, and they reproduce quite naturally the paucity of BDs in close orbits. Another possible formation mechanism is disk fragmentation. Numerical simulations \citep[e.g.][]{thies2010} have shown that numerous BDs can be formed in the outer regions of massive circumstellar disks. The mass spectrum of objects formed around a 0.7 \msol\ star has a maximum around the mass of \205 b \citep[Fig.~4 of][]{stamatellos2009}, so the measured mass is not only understood by but also expected from theoretical considerations. However, both processes form BDs no closer than 10 AU from the star, and therefore a mechanism is needed to migrate the newly-formed object to the inner system.

Scattering by other objects formed around the same star \citep[e.g.][]{fordrasio2008} or the Kozai mechanism \citep{kozai1962}  can bring the object to a closer orbit, possibly resulting in an eccentric and non-aligned orbit. Subsequent tidal dissipation is expected to have rapidly circularized the system \citep[e.g.][]{matsumura2008}. In these cases, the presence of at least one additional object in the system would be expected, but it is also reasonable to assume that these additional bodies are either very far out or have been ejected. Unfortunately, the time span of our current RV data is not sufficient to probe these possibilities.

\begin{figure}[t]
\begin{center}
\includegraphics[width = \columnwidth]{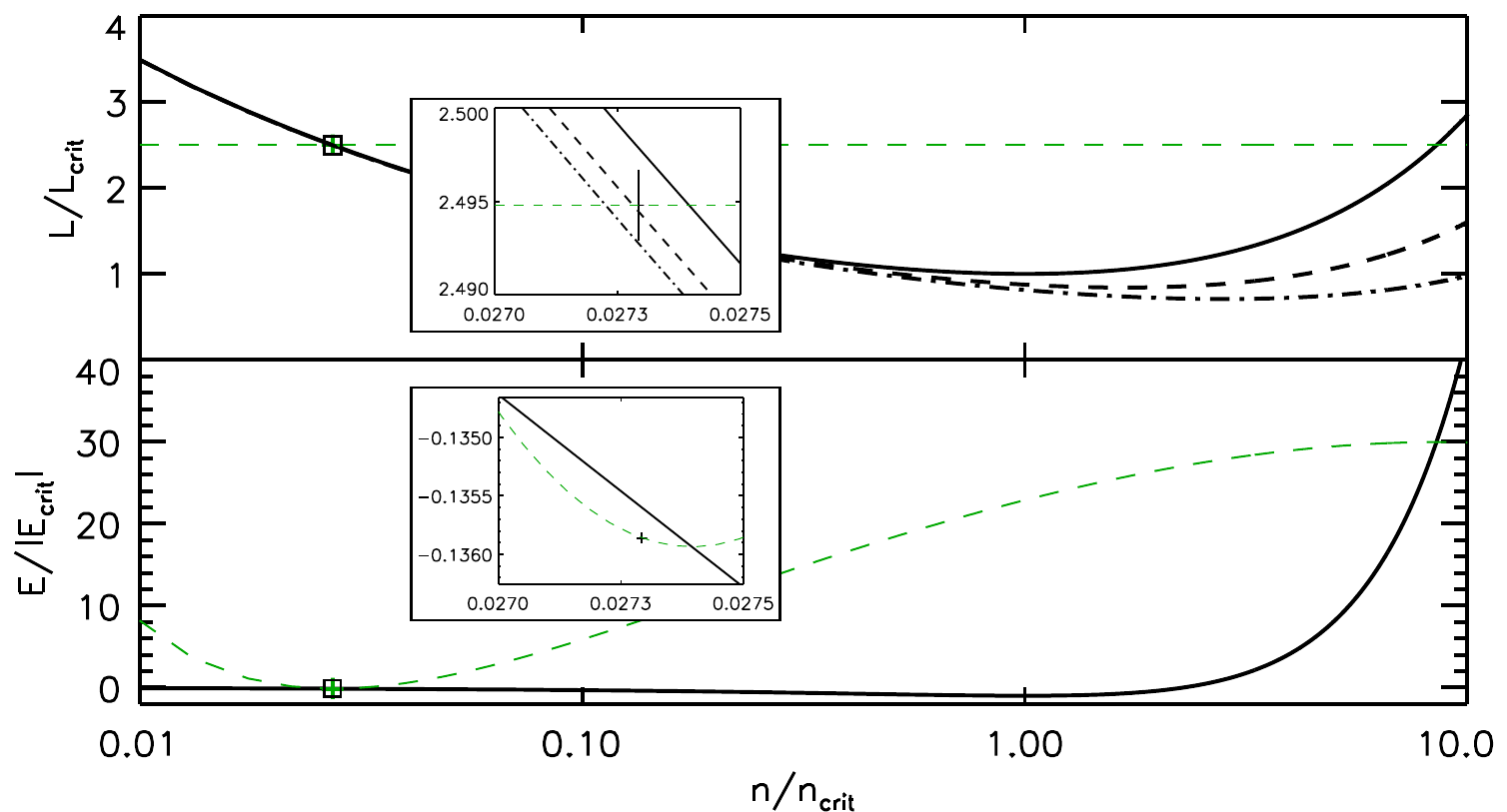}
\caption{Total angular momentum (\emph{top}) and energy (\emph{bottom}) of the system as a function of angular orbital frequency $n = 2\pi/P$, scaled to the critical values (see text). The solid curve is the corotation locus ($n = \Omega_s = 2\pi/P_\mathrm{rot}$) for zero eccentricity, and the dashed and dot-dashed curves are the $n = 2\Omega_s$ and $n = 4\Omega_s$ loci, respectively. The green dashed curve is the constant total angular momentum locus. The position of the \205 system as of today is marked by the square that is zoomed in the insets.\label{fig.tides}}
\end{center}
\end{figure}

The fact that the orbit is nearly circular suggests that the tidal interactions are strong in the system, especially considering the spectral type of the star and the relatively high mass ratio. In fact, as mentioned in the introduction, close massive companions to late-type stars are not expected to survive engulfment.  However, by taking the stellar spin rate from the spectroscopic $v \sin i_*$ and assuming that the rotation of the BD is synchronized with its orbital period\footnote{The characteristic synchronization timescale is $\sim 500$ Myr, which is short compared to the age of the system.}, it can be seen that the system is able to evolve toward a stable configuration (Fig.~\ref{fig.tides}) . Indeed, the total angular momentum of the system is higher than the critical value \citep{hut1980} while the orbital frequency is higher than the stellar rotation frequency. Therefore, tidal evolution will lead \205 b to migrate toward the star but engulfment will be prevented because the system will reach a stable equilibrium state at corotation. This can be clearly seen from the bottom panel of Fig.~\ref{fig.tides}, which shows an energy minimum of the constant angular momentum curve at corotation.

This picture is complicated by the magnetic braking of the star, which leads to a loss of total angular momentum. But even considering a Skumanich-type law for the loss of angular momentum \citep{spada2011}, and considering the extreme case where the escaped angular momentum is drawn entirely from the \emph{orbital} angular momentum at constant stellar spin, the system is expected to be stable on the scale of thousands of gigayears \citep{hut1980}. Furthermore, if \205 were a more massive G-type star with a five-day rotational period, the system would still be stable over more than 10 Gyr.

This shows that stars with convective envelopes are capable of harboring massive companions in close orbits. Therefore, the lack of  detections around late-type stars cannot be explained by their being doomed to fall into their stars. On the other hand, an inefficient formation process, selection biases, or insufficient statistics might be invoked.  Whichever is the case, \205 b will prove to be of great importance in constraining formation and evolution theories of brown dwarfs and massive extrasolar planets.

\input{tableParams_onlyBestSolution_v2.tex}

\begin{acknowledgements}
We thank the staff at Haute-Provence Observatory. We acknowledge the PNP of CNRS/INSU, and the French ANR  for their support. This publication makes use of data products from the Wide-field Infrared Survey Explorer. RFD is supported by CNES. 
\end{acknowledgements}

\bibliographystyle{aa} 
\bibliography{biblio_iafe}

\end{document}

%% file: tableParams_onlyBestSolution_v2.tex
\setlength{\tabcolsep}{1pt}
\begin{table}[t]
\caption{Orbital and physical parameters. \label{table.Params}}
\vspace{-0.1cm}
\begin{tabular}{p{0.6\columnwidth}rcl}
\hline
\hline
\noalign{\smallskip}
Period $P$\tablefootmark{*}						[days]				& 11.7201248 &$\pm$& 2.1e-06\\
Midtransit time $T_c$\tablefootmark{*}				[BJD]\tablefootmark{a}	& 975.17325 &$\pm$& 1.2e-04\\
Eccentricity $e$\tablefootmark{*}				 						&\multicolumn{3}{c}{$< 0.031$\tablefootmark{b}}\\
%\noalign{\smallskip}
Argument of periastron $\omega$\tablefootmark{*}		[deg] 				& 263 &$^+_-$&$^{61}_{230}$ \\
%\noalign{\smallskip}		
Inclination $i$\tablefootmark{*}						 [deg]				& 88.456 &$\pm$& 0.055	\\
%\noalign{\smallskip}
\hline
\noalign{\smallskip}
%\noalign{\bigskip}
Radial-velocity amplitude $K$\tablefootmark{*}		 	[km s$^{-1}$]			& 3.732 &$\pm$& 0.039 	\\		
Radius ratio $k  = R_b/R_s$\tablefootmark{*} 								& 0.09849 &$\pm$& 0.00049	\\		
Mass ratio $q  = M_b/M_s$\tablefootmark{*}	 								&  [0.0 &-& 0.02]\tablefootmark{c}	\\		
Center-of-mass velocity $\gamma$\tablefootmark{*} 		 [km s$^{-1}$] 			& 15.057 &$\pm$& 0.026\\		
\noalign{\smallskip}
Semi-major axis					[AU]					& 0.0987 	&$\pm$& 0.0013	\\
Transit duration					[hours]				&  3.07 	&$\pm$& 0.15	 \\
%\noalign{\smallskip}
%\multicolumn{3}{l}{\it Stellar Parameters}\\
\hline
\noalign{\smallskip}
%\noalign{\bigskip}
Stellar density $\rho_s$\tablefootmark{*}				 [$\rho_\odot$]			& 1.550 &$\pm$& 0.073	\\
Effective temperature $T_\mathrm{eff}$\tablefootmark{*}	 [K]					& 5237 &$\pm$& 60 \\
Metallicity [Fe/H]\tablefootmark{*}					 [dex]				& 0.14  &$\pm$&0.12\\		
Distance $d$\tablefootmark{*}						 [pc]					& 585  	&$\pm$& 16	\\
E(B-V)\tablefootmark{*}							 [mag]				& 0.040 	&$\pm$& 0.023	\\	\noalign{\smallskip}
Rotational velocity $v\sin i_*$	[\kms]					& 2.0 	&$\pm$& 1.0		\\
Rotation period [d]										& 40.99	&$\pm$& 0.50		\\
\noalign{\smallskip}
Stellar mass $M_s$					 [\msol]				& 0.925 &$\pm$& 0.033\\
Stellar radius $R_s$				 [\RS]				& 0.841	 &$\pm$& 0.020	\\
%\noalign{\smallskip}
%Surface Gravity $\log g$ 				 [cm s$^{-2}$] 			& 4.555 &$\pm$& 0.014\\
%
%Luminosity $\log(L_s/L_\odot)$			 					& -0.321 &$\pm$& 0.033	\\
%
Age								 [10$^9$ yr]			& [0.4	&-&	8.3]\tablefootmark{d}\\%\multicolumn{3}{c}{[0.4 - 8.3]\tablefootmark{a}}\\

%\noalign{\smallskip}
%\multicolumn{3}{l}{\it Companion Parameters}\\
%\hline
\hline
\noalign{\smallskip}
Companion mass $M_b$						 [\MJ]					& 39.9 	&$\pm$& 1.0	\\
Companion radius $R_b$						 [\RJ]					& 0.807	&$\pm$& 0.022		\\
Companion density $\rho_b$					 [$\rho_\mathrm{Jup}$]	& 75.6	&$\pm$& 5.2		\\		
Equilibrium temperature $T_\mathrm{eq}	$ [K]					& 737	&$\pm$& 31		\\		
%\noalign{\smallskip}	
\hline
\hline
\end{tabular}
\vspace{-0.1cm}
\tablefoot{
\tablefoottext{*}{Fitted parameters in the MCMC algorithm}
\tablefoottext{a}{BJD\_UTC  - 2,454,000}
\tablefoottext{b}{99\% upper limit}
\tablefoottext{c}{uniform distribution; not constrained.}
\tablefoottext{d}{99\% confidence interval}
}

\end{table}